\documentclass[letterpaper,10pt,twocolumn]{article}
\usepackage{amssymb,amsmath,graphicx,color,ulem}
\usepackage[superscript,biblabel]{cite}
\usepackage[margin=2cm]{geometry}
\usepackage{authblk}
\usepackage[symbol*]{footmisc}
\newif\ifcomments

\commentstrue

\ifcomments
\newcommand{\tjcom}[1]{\textcolor[rgb]{0,0,0.5}{[{\scriptsize TJ-com:}{\small #1}]}}
\newcommand{\shcom}[1]{\textcolor[rgb]{0.5,0,0}{[{\footnotesize SH-com:}#1]}}
\newcommand{\fdcom}[1]{\textcolor[rgb]{0,0.5,0}{[{\footnotesize FD-com:}#1]}}
\newcommand{\tjadd}[1]{\textcolor[rgb]{0,0,0.5}{[add: #1]}}
\newcommand{\shadd}[1]{\textcolor[rgb]{0.5,0,0}{[add: #1]}}
\newcommand{\fdadd}[1]{\textcolor[rgb]{0,0.5,0}{[add: #1]}}
\newcommand{\tjdel}[1]{\textcolor[rgb]{0,0,0.5}{[del: \sout{#1}]}}
\newcommand{\shdel}[1]{\textcolor[rgb]{0.5,0,0}{[del: \sout{#1}]}}
\newcommand{\fddel}[1]{\textcolor[rgb]{0,0.5,0}{[del: \sout{#1}]}}
\else
\newcommand{\tjcom}[1]{}
\newcommand{\shcom}[1]{}
\newcommand{\fdcom}[1]{}
\newcommand{\tjadd}[1]{}
\newcommand{\shadd}[1]{}
\newcommand{\fdadd}[1]{}
\newcommand{\tjdel}[1]{}
\newcommand{\shdel}[1]{}
\newcommand{\fddel}[1]{}
\fi

\newcommand{\bq}{\begin{equation}}
\newcommand{\eq}{\end{equation}}
\newcommand{\er}[1]{Eq.\ \eqref{#1}}

\newcommand{\vsc}{V_\text{sc}}
\newcommand{\vh}{V_\text{H}}

\newcommand{\nss}{N_\text{ss}}
\newcommand{\ess}{E_\text{ss}}
\newcommand{\esso}{E_{\text{ss}}^{\text{o}}}
\newcommand{\besso}{\bar{E}_\text{ss}^\text{o}}

\newcommand{\ecat}{E_\text{cat}}

\newcommand{\esol}{E_\text{sol}}

\newcommand{\nb}{\bar{n}}
\newcommand{\pb}{\bar{p}}
\newcommand{\vbsol}{\text{vb,sol}}
\newcommand{\cbsol}{\text{cb,sol}}

\newcommand{\vbcat}{\text{vb,cat}}
\newcommand{\cbcat}{\text{cb,cat}}

\newcommand{\catsol}{\text{cat,sol}}
\newcommand{\vbss}{\text{vb,ss}}
\newcommand{\cbss}{\text{cb,ss}}
\newcommand{\sscat}{\text{ss,cat}}
\newcommand{\sssol}{\text{ss,sol}}
\newcommand{\jo}{\bar{J}}

\newcommand{\e}{\epsilon}

\newcommand{\vap}{V_\text{app}}
\newcommand{\bvh}{\bar{V}_\text{H}}
\newcommand{\jocs}{\tilde{J}_\text{cat,sol}}
\newcommand{\voc}{V_\text{oc}}
\newcommand{\joss}{\tilde{J}_\text{ss,sol}}
\newcommand{\vho}{\bar{V}_\text{H}}
\newcommand{\ch}{C_\text{H}}

\makeatletter
\let\@fnsymbol\@arabic
\makeatother
\newcommand\blfootnote[1]{%
  \begingroup
  \renewcommand\thefootnote{}\footnote{#1}%
  \addtocounter{footnote}{-1}%
  \endgroup
}

\title{The role of surface states in electrocatalyst-modified semiconductor photoelectrodes: Theory and simulations}
\author{Thomas J. Mills,\thanks{published posthumously} \ Forrest A.L. Laskowski, Christian Dette, Michael R. Nellist, Fuding Lin,} \author{Shannon W. Boettcher*}
\affil{\it{Department of Chemistry and Biochemistry, University of Oregon, Eugene, Orgeon 97403, USA}}
\begin{document}
\date{}
\maketitle

\section{Introduction}

In the last several years there has been a wealth of studies performed to clarify the mechanism of the oxygen evolution reaction (OER) on semiconducting photoelectrodes and how it is modified by the addition of thin layers of electrocatalysts or other materials.\cite{barroso11,barroso12} Similar trends are seen with several semiconductor materials, such as Fe$_2$O$_3$, TiO$_2$, BiVO$_4$, and WO$_3$, with overlayers based on oxides of Co, Ni, Fe, Ga, and Al.\cite{ding13,choi13,lichterman13,long08,seabold11,zhong11,kim14,liao12,mcdonald11,spray11} The addition of a thin oxide overlayer in many cases cathodically shifts the potential of photocurrent onset and/or increases the maximum photocurrent, leading to greater collection efficiencies.\cite{hisatomi11,zhong09} In order to take advantage of this improved efficiency it is therefore important to understand its origin and the factors that would lead to its optimization. \blfootnote{*swb@uoregon.edu}

A number of related mechanisms have been proposed to explain the role of the oxide overlayer, each supported by data from a range of analysis methods.\cite{gamelin12} It is generally agreed that recombination is the main loss pathway and that the overlayers function to suppress surface recombination, but there are several ways in which they can do this. Transient absorption spectroscopy (TAS) has demonstrated that water oxidation requires the existence of long-lived photogenerated holes in bare Fe$_2$O$_3$, TiO$_2$, and WO$_3$.\cite{cowan11,cowan10,pendlebury11,pesci11,tang08} A more positive bias depletes the space charge region of electrons, reducing recombination and allowing for the long-lived hole species to oxidize water. It has been found that Co and Ga oxide overlayers cathodically shift the potentials required for the existence of these species. Such a phenomenon could be explained by the semiconductor-oxide interface forming an n-p heterojunction, by the oxide depleting electron density from the semiconductor space charge region, by the oxide separating charge, or by the oxide increasing the band bending in the semiconductor; these explanations are not all entirely distinct from one another, but they all imply that the oxide plays a non-catalytic role in the OER, increasing the overall rate by modifying the nature of the semiconductor-electrolyte interface.

Other evidence points towards a catalytic role particularly of Co oxide layers. It has been suggested that Co oxide layers function to increase the kinetics of the OER, removing the bottleneck of charge transfer to solution, leaving the system limited by the rate of hole collection from the semiconductor.\cite{klahr12} In this view the overlayer does not modify the energetics of the interface, but functions in a traditionally catalytic way by speeding up the interfacial charge transfer. By moving positive charge out of the semiconductor and into the catalyst and solution, this mechanism also has the ultimate effect of decreasing surface recombination. Another view suggests that the overlayers passivate semiconductor surface states, thus suppressing surface recombination directly rather than through modifying the nature of the space charge region.\cite{le11,liang14}

Many of these proposed mechanisms are rather similar and to some degree indistinguishable from each other, although they all rely on mitigating surface recombination. Perhaps the largest physical distinction that can be measured is whether or not this reduction is due to a change in the space charge layer, or due to some modification of the semiconductor-electrolyte interface produced by the oxide layer. It has been verified by Mott-Schottky analysis that the flat-band potential is not appreciably modified by the addition of an overlayer;\cite{chou13} this is consistent with the use of very thin layers that cannot store enough charge to do so.

To probe the nature of the kinetics and of the space charge region, a few models have been proposed to interpret the results of spectroscopy techniques, mainly electrochemical impedance spectroscopy (EIS) and intensity modulated photocurrent spectroscopy (IMPS).\cite{chou13,leng05,ponomarev95,wilson77} These techniques provide information about the kinetics of charge transfer and recombination, but the analysis is somewhat involved and its interpretation is not unique.\cite{vanmaekelber97,peter12,peter97,upul11} Equivalent circuit models have often been used, but more in-depth analyses using kinetic mathematical models have shown that interpretation of spectra is not so straightforward and the use of equivalent circuit models may be misleading - in other words, the physical processes involved do not correspond neatly to simple capacitors and resistors. The use of these kinetic models is somewhat limited though by the complicated expressions relating the spectra to the kinetic parameters of the physical system.

We aim here to formalize and quantify arguments that have been made heretofore mainly qualitatively. It is not difficult to write down simple expressions that describe the kinetic and electrostatic processes occuring in these systems, but it is difficult to concisely describe the predictions of a full kinetic model. There are several main processes that need to be considered: transport of photogenerated holes to the space charge region, filling and emptying of surface states, charge transfer between semiconductor, surface states, catalyst, and solution, and the division of potential drops between the space charge and Helmholtz regions, which is modified by the filling of the surface states. Kinetic models describing most of these processes have been used in deriving the impedance response, although these have not included an explicit overlayer.

Instead of describing the effect of these processes on the impedance spectra or presenting complicated mathematical results, we will use kinetic model equations to describe the steady-state $J(V)$ response of an idealized combined semiconductor-surface state-catalyst-solution system. In this way we can directly show what changes in the various processes do to the collection efficiency and where these effects come from by explicitly calculating the electrostatic and electrochemical potentials of each component of the system.

For example, while the effect of surface state filling on the potential drops has been invoked in a qualitative way, there a number of subtleties that arise because of the coupled nature of the kinetics and the electrostatics. It is known that transfer can occur from semiconductor surface states to solution, and in this way surface states can function not only as recombination centers, but also conduits by which charge can be transferred indirectly from semiconductor to solution (surface state-mediated transfer); they can in fact in certain circumstances increase transfer despite recombination. Furthermore, it has been noted that the kinetics of charge transfer to solution can affect the electrostatics, for if charge is transferred out of the surface states more quickly, their filling will go down, leaving more potential to drop across the space charge region.

This latter effect means that an overlayer that functions catalytically can also affect the band bending and space charge regions by changing the filling of the surface states. This is particularly important when there is a high density of surface states (as there frequently are in the semiconductors typically used), where Fermi level pinning is important; in this case, a catalyst that shuttles charge into the solution not only increases the kinetics at the interface, but delays the onset of pinning, while also reducing recombination by reducing the surface state filling. In this sense, a catalytic overlayer is indistinguishable from one which decreases recombination by altering the space charge region - the two effects are closely coupled, but it could be asked, which is the cause, and which is the effect?

Our previous work has shown how the addition of ion-permeable catalysts to the surface of semiconductors affects the current/voltage response using full numerical simulations of the semiconductor transport and interfacial transfer equations.\cite{mills14} Here we extend these results by first showing how the semiconductor charge transport can be analytically approximated, removing the need for the numerical solution of differential equations. We then include surface states and calculate the surface state filling and the potential drop caused by surface charge. This allows us to obtain a relatively simple approximate analytical model that can easily be numerically solved to yield the potentials of each part of the system and the $J(V)$ response. Using this we can demonstrate in a quantitative way what effect each different possible mechanism of catalysis or reduction of recombination the overlayers can have on the collection efficiency of the combined system.

\section{Model}

\begin{figure}\begin{center}\includegraphics[scale=.9]{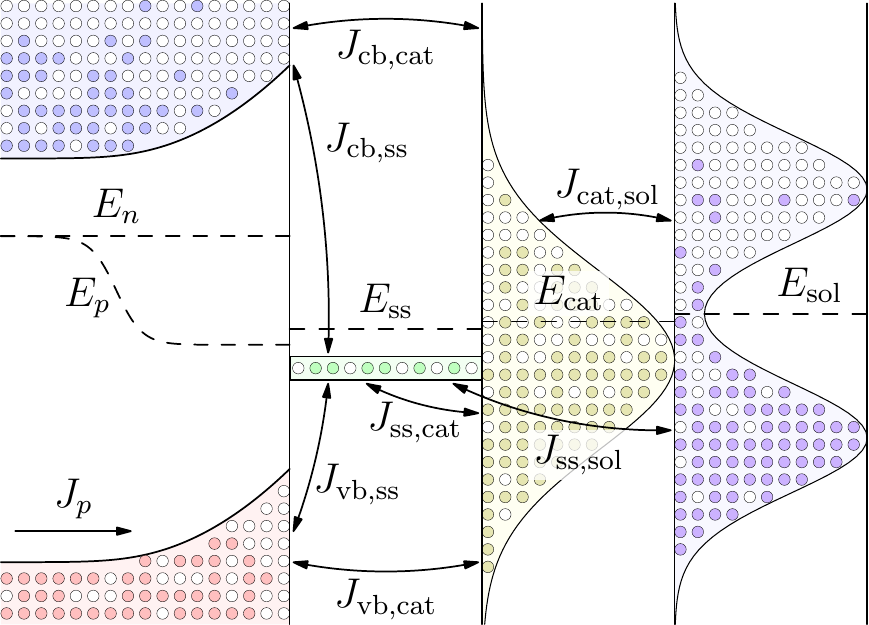} \caption{Diagram of the investigated model including the semiconductor (sc), surface states (ss), catalyst (cat) and the solution (sol).} \label{sysdiag}\end{center}\end{figure}

\subsection{Terminology: catalyst sites and surface states}

From a modeling perspective, surface states and catalyst sites are very similar; both can react with electrons and holes in the semiconductor, functioning as recombination centers, and both can transfer charge to the solution, acting as OER sites. For example, it has been suggested that Fe$_2$O$_3$ oxidizes water through high oxidation state Fe centers on the surface \cite{upul11,peter12}; in this case, the surface states act both as recombination and oxidation centers. Overlayers of Al and Ga oxides increase efficiency, but presumably do not transfer charge to the solution; it has thus been suggested that these overlayers do not act catalytically, but instead modify the surface in a way that decreases recombination \cite{barroso12,hisatomi11,le11}.

The physical distinction between semiconductor surface states and additional overlayers is clear, but both may function as recombination and/or oxidation sites. In this work we investigate only catalytic overlayers, so we will use the term ``catalyst'' to refer to these overlayers, bearing in mind that surface states can also act catalytically. Hence, the main distinguishing feature of surface states and catalyst sites is that surface states are an intrinsic part of the semiconductor material, whereas catalysts form a separate phase attached to the surface, and are often thin, porous, ion-permeable overlayers \cite{klahr12,lin14}.

We will assume in this work that the catalysts are ion-permeable. Because of this ion permeability, ions from the solution can approach the semiconductor surface and screen the electric field there, so that the Helmholtz layer and its potential drop occur at the semiconductor surface rather than the catalyst-solution interface. In other words, catalyst sites can be screened by solution ions, whereas surface states cannot. Surface states often occur in high enough concentration that charging them can affect the Helmholtz potential, thus also altering the potential drop across the semiconductor depletion region and in turn the surface electron and hole concentrations.

In this work we exclude transfer directly from semiconductor to solution; this is unlikely to be a significant source of current since the OER is a four-electron process \cite{klahr12}. For clarity we will assume that the surface states are all at a single energy $\esso$, as is done in the traditional Shockley-Read-Hall surface recombination model. It has been shown that the existence of a continuum of surface states is required to explain photocurrent at a TiO$_2$-water interface \cite{wilson77}, but a single-energy model will suffice as a first approximation. We will also treat a one-electron redox reaction instead of the actual four-electron OER to avoid the complications of multi-step reactions. Only positively charged surface states are considered since these are the ones that will play a role in the oxidation process, although negatively charged states can affect both surface recombination and the Helmholtz potential.

\subsection{Notation, variables, and parameters}

The semiconductor surface is at $x=0$, and the semiconductor extends in the negative $x$ direction, so that positive currents represent net current into the solution. The total semiconductor width is assumed to be large relative to the depletion width so that the bulk semiconductor corresponds to $x \sim -\infty$. Electron and hole densities in the semiconductor are labeled $n$ and $p$, respectively, with a subscript $s$ indicating the value at the surface ($x=0$). Current density is labeled $J$ and has two subscripts corresponding to transfer between two subsystems, except where indicated. The subsystems are labeled vb (valence band), cb (conduction band), sc (semiconductor), ss (surface states), cat (catalyst), and sol (solution). The electrostatic potential is labeled $\phi$ and the total electrostatic potential drop across the entire system is $V$. Electrochemical potentials (quasi-Fermi levels) of the subsystems are labeled $E$ with a subscript indicating the subsystem, with the exception of $E_C, E_V$ which represent the conduction and valence band edge energies, and $E_n, E_p$ which represent the electron and hole quasi-Fermi levels. The energy level (standard potential) of the surface states is $\esso$. Overbars (e.g.\ $\bar{n}$) indicate equilibrium quantities; in the case of current densities, overbars indicate exchange currents (i.e.\ the unidirectional equilibrium currents rather than the total equilibrium current, which is zero). The equilibrium electron $\nb$ and hole $\pb$ concentrations without subscripts indicate the equilibrium bulk concentrations. Standard symbols are used for physical constants: $k$, $T$, $q$, $\varepsilon$, indicate the Boltzmann constant, absolute temperature, magnitude of elementary charge, vacuum permittivity... Material parameters are the hole diffusion coefficient $D_p$, effective density of states constants for the conduction and valence bands $N_C$ and $N_V$, semiconductor absorption coefficient $\alpha$, semiconductor diffusion length $\delta$... 
A few quantities are computed from the parameters: the semiconductor Debye length $\lambda \equiv \sqrt{\varepsilon \varepsilon_0 kT /q^2 N_d}$, where $\varepsilon$ is the dielectric constant and $N_d$ is the donor density ($N_d \approx \nb$ in an \textit{n}-type semiconductor); and the semiconductor hole diffusion length $\delta \equiv \sqrt{D_p k_R/\nb}$, where $k_R$ is the second-order recombination rate constant.

All quantities are written in physical units except for energies and potentials, which are treated as unitless quantities that have been reduced by the thermal energy $kT$ (for energies) or the thermal voltage $kT/q$ (for potentials). Because of this variable reduction, there are many equations in this work that appear to treat potentials and energies as though they have the same units - the reader should keep in mind that in physical units there is a proportional factor of $q$ in such equations.

All energies are referenced to the solution potential $E_\text{sol} \equiv 0 $, and the sign convention is chosen to produce an electron energy scale, so that more negative potentials are more oxidizing (the opposite sign convention from that commonly used in electrochemistry). The electrostatic potential is referenced to the bulk semiconductor, $\phi(-\infty) \equiv 0$, so that the electrostatic potential in the solution is $\phi_\text{sol} = -V$. Note that energies and electric potentials are referenced to opposite ends of the system ($x=\infty$ for energies, $x=-\infty$ for potentials). Relating energies in combined solid state-electrochemical systems can be challenging because of the different energy scales used; for a thorough exposition of the relation of these scales and understanding energy diagrams, we refer the reader to the work of Bisquert\cite{Bisquert14}, which uses similar notation to ours. 

In the isolated semiconductor at equilibrium, $E_n=E_p=\bar{E}_\text{sc}$, $E_C=\bar{E}_C$, and $E_V=\bar{E}_V$. The carrier concentrations are given by
\bq \bar{n} = N_C e^{\bar{E}_\text{sc} - \bar{E}_C } \qquad \bar{p} = N_V e^{\bar{E}_V-\bar{E}_\text{sc}} \eq
After equilibration with the rest of the system, the band edges shift according to
\bq E_C = \bar{E}_C - (\phi-\phi_\text{sol}) \qquad E_V = \bar{E}_V - (\phi-\phi_\text{sol}) \eq
The bulk concentrations remain at their pre-equilibrium values. The semiconductor concentrations and quasi-Fermi levels are then related by
\bq n = \nb e^{-E_n + (\phi - \phi_\text{sol})} \qquad p = \pb e^{E_p - (\phi - \phi_\text{sol})} \eq
The total electrostatic potential drop at equilibrium is $\bar{V} \equiv \bar{E}_\text{sc} - \esol$. The applied bias $\vap \equiv V - \bar{V}$ is the externally applied change in potential from its equilibrium value. Note that the applied bias also shifts the electrochemical potentials, so that $\vap = E_\text{sc} - E_\text{sol}$.

The Helmholtz potential drop is defined as the difference in potential between the semiconductor surface and the solution, $\vh \equiv \phi_s - \phi_\text{sol}$, where $\phi_s$ is the potential at the semiconductor surface ($x=0$). The surface electron and hole concentrations are therefore given by
\bq n_s = \nb e^{-E_{n,s} + \vh} \qquad p_s = \pb e^{E_{p,s} - \vh} \eq

\subsection{Electrostatic potential} \label{epot}

The electrostatic potential drops across the semiconductor depletion region $\vsc \equiv -\phi_s$ and the Helmholtz layer drop $\vh \equiv \phi_s - \phi_\text{sol}$ are determined by electroneutrality, i.e.\ by equality of charge on either side of the semiconductor-solution interface \cite{bockris73}:
\bq q_\text{sc} + q_\text{ss} = q_\text{H} , \label{potqeq} \eq
where $q_\text{sc}$ is the excess charge in the depletion region of the semiconductor, $q_\text{ss}$ is that in the semiconductor surface states, and $q_\text{H}$ is that in the Helmholtz layer. The sum of the potential drops should equal the total potential drop across the entire system,
\bq \vsc + \vh = V .\eq
With a small amount of surface charge $q_\text{ss}$, ions in the Helmholtz layer compensate the charge in the depletion region $q_\text{sc}$; because the electrolyte concentration is generally much higher than the dopant density, the Helmholtz potential drop is typically quite small in the absence of surface charge. When there is enough charge in the surface states ($q_\text{ss} \gg q_\text{sc}$), more ions will be needed to balance the charge, potentially leading to an increased Helmholtz potential drop. Since the catalyst charge is distributed throughout the ion-permeable catalyst layer, ions from the solution can balance the catalyst charge outside of the Helmholtz layer, so we can assume that the catalyst charge does not enter into the interfacial charge neutrality condition [\er{potqeq}].

An important consequence of the Helmholtz potential is that it shifts the energy of the surface states $\esso$ relative to the solution. Denoting by $\besso$ the value of $\esso$ in the absence of a Helmholtz potential, we have $\esso = \besso - \vh$.

Using the depletion layer approximation, the charge in the semiconductor depletion region, assuming $\vsc>0$, is
\bq q_\text{sc} \approx q N_d \lambda \sqrt{2 \vsc} \eq
The charge in the surface states is determined by the surface state electrochemical potential $\ess$ and the surface state DOS $g_\text{ss}$. Since we have chosen to use a single energy level surface state model, $g_\text{ss}(\e) = \nss \delta(\e-\esso)$, and
\bq q_\text{ss} = q 
\int g_\text{ss}(\e) f(\ess-\e)\, d\e = \frac{q\nss}{1+e^{\ess-\besso+\vh}} . \eq
The Helmholtz region is essentially a capacitor in which one electrode is the semiconductor surface and the other is the layer of ions in the Helmholtz plane, with a neutral region between them, so
\bq q_\text{H} = C_\text{H} \vh , \label{potqh} \eq
where $C_\text{H}$ is the capacitance of this region. $C_\text{H}$ is determined by the dielectric constant and width of the region, ... \cite{bockris73}.

\begin{figure}\begin{center}\includegraphics[scale=0.8]{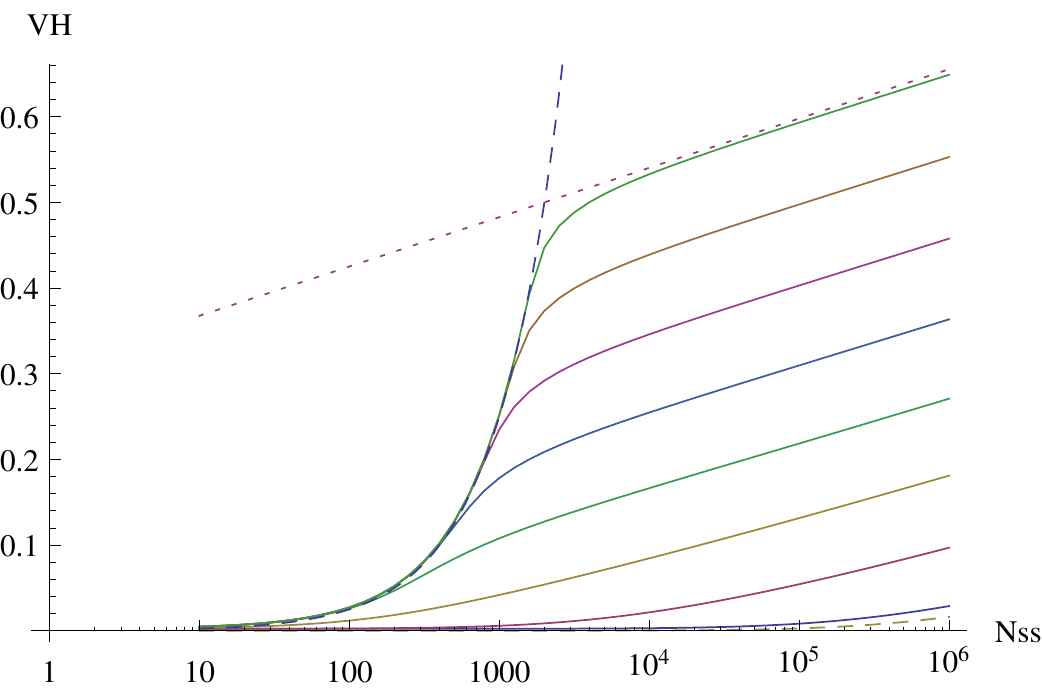} \caption{The equilibrium Helmholtz potential $\bvh$ as a function of $\nss$ for $\besso = -0.2, -0.1, ... 0.5$ V.} \label{phirat}\end{center}\end{figure}

Equations \eqref{potqeq} through \eqref{potqh} determine the division of the total electrostatic potential $V$ into $\vsc$ and $\vh$ in terms of the parameters and the surface state potential $\ess$. At equilibrium $\ess=0$, and the equilibrium potential drops $\bar{V}_\text{sc}$ and $\bar{V}_\text{H}$ are constants determined by the parameters. Figure \ref{phirat} shows $\bar{V}_\text{H}$ as a function of $\nss$ for $\besso = -0.25$ to $0.5$ V. When $\nss$ is small, $\bvh$ is approximately proportial to $\nss$,
\bq \bvh \sim kT \frac{\nss}{C_\text{H}} \quad (\nss \sim 0) . \eq
This limiting curve is shown in dashed lines in Figure \ref{phirat}.

When $\nss$ is large, $\bvh$ surpasses $\besso$ - the surface state energy has been shifted all the way past the solution potential - and its dependence on $\nss$ becomes much weaker,
\bq \bvh \sim \besso + \ln\left( kT \frac{\nss\besso}{C_\text{H}} \right) \quad (\nss \sim \infty) . \eq
This limit is shown in dotted lines for $\besso = 0.5\ \text{V}$. In this case, there are enough surface states that, if they remained filled, would produce a very large potential; the system acts to bring the surface state energy $\esso$ down far enough to unfill enough of them to minimize the potential. Under applied bias, this principal continues to work and acts to keep the Helmholtz potential roughly constant; this is known as the \textit{Fermi level pinning} regime. Since this effect occurs when $\bvh$ reaches $\besso$, we will be in the pinning regime approximately when
\bq \nss >\frac{C_\text{H} \besso} { kT } \eq
Note that the above analysis requires that $\besso > 0$ and that the contribution from the depletion layer can be neglected. See the SI for more solution information.

\subsection{Semiconductor}

\begin{figure}\begin{center}\includegraphics[scale=.95]{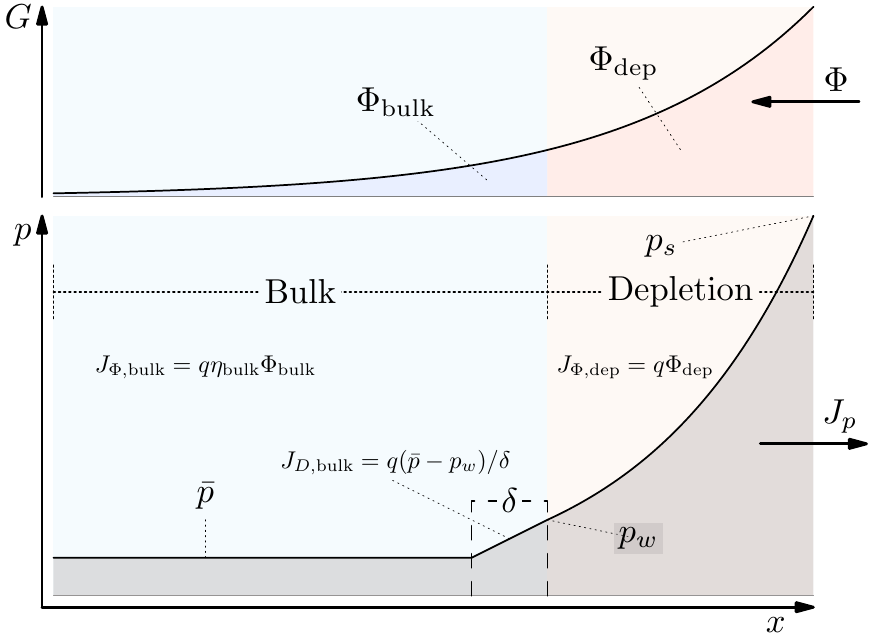} \caption{Hole concentration profile and the currents generated by the generalized G\"artner model.} \label{holediag}\end{center}\end{figure}

It is commonly assumed \cite{peter97,peter12} that the surface electrons are at quasiequilibrium with the bulk ($E_n \approx 0$), so that
\bq n_s = \nb e^{-\vsc} \label{nsqeq} \eq
and that the hole current $J_p$ is equal to the G\"artner current \{cite G\"artner\}
\bq J_G = \Phi \left( 1 - \frac{e^{-\alpha w}}{1+\alpha \delta} \right) + q D_p \frac{\pb}{\delta} \label{jg} \eq
Our previous work \cite{mills14} explicitly solved the semiconductor transport and continuity equations numerically to obtain the surface hole and electron densities $p_s$ and $n_s$. We found that while \er{nsqeq} is a good approximation, setting $J_p = J_G$ is not always appropriate, particularly with ion-permeable catalysts. This is because, when the catalyst becomes oxidized, there are fewer available neutral sites to oxidize, requiring more holes to oxidize it further, so that a larger value of $p_s$ is needed to sustain the current. The semiconductor may not be able to provide enough holes for the current to reach $J_G$, in which case the forward current is limited by transport of holes to the interface. This effect is more pronounced in systems with ion-permeable catalysts, because the catalyst can become highly oxidized at much lower biases than in systems with impermeable catalysts \cite{mills14}. The G\"artner model neglects the behavior of the holes in the depletion region, assuming perfect hole conductivity with no limitation. It has been recognized that the G\"artner model is insufficient when the reaction kinetics are slow due to buildup of minority carriers in the depletion region \cite{el82,albery83, peter14}.

In order to treat the hole transport problem more realistically we require a generalization of the G\"artner model that accounts for the hole transport limitation; without this limitation $E_p$ could in principle decrease indefinitely, leading to unrealistically high values of $p_s$. This effect has been included in other treatments \cite{reiss78}, although usually such treatments aim to obtain a complete explicit solution for $J(V)$. Our model will be phrased implicitly in terms of $p_s$ so that it can be coupled to boundary conditions later.

There are two main effects of the buildup of minority carriers in the depletion region due to slow kinetics: a large diffusional back-current due to the high concentration gradient, and increased recombination in the depletion region. The latter effect has been treated before \cite{albery81,albery83,reichman80} but the analysis is rather involved; here we use a simple approximation of the hole concentration profile to treat depletion region recombination and discuss the errors in this approximation in the SI (see Section \ref{gart}).

\subsubsection{The Generalized G\"artner Model}\label{gart}

This derivation closely follows the original one of G\"artner \cite{gartner59}. The two generalizations we make are (1) allowing for a non-zero value of the hole concentration at the edge of the depletion region (corresponding to relaxing G\"artner's original assumption of fast surface kinetics), and (2) allowing for recombination in the depletion region, which requires approximating the hole concentration profile in this region. The derivation is given in full to show how the generalizations fit naturally into the original treatment. Extensive analyses of the minority carrier profiles and depletion region recombination have been conducted by Albery et.\ al.\ \cite{albery81,albery83}; our treatment is simpler, but a basic comparison of the errors in the two methods are included in the SI. Our method and results are closely related to those of El Guibaly et.\ al.\ \cite{el82} but we use simple second-order recombination rather than trap-mediated recombination.

Figure \ref{holediag} shows quantities relevant to the derivation of the generalized G\"artner model. The incoming photon flux $\Phi$ is split into a portion that is absorbed in the depletion region, $\Phi_\text{dep}$, and in the bulk, $\Phi_\text{bulk}$, so that $\Phi = \Phi_\text{dep} + \Phi_\text{bulk}$. The semiconductor interface is at $x=0$ and the inside of the depletion region is at $x=-w$, where $w$ is the depletion region width. The generated carriers per unit time is
\bq G(x) \equiv \Phi \alpha e^{\alpha x} \eq
Recombination is assumed to follow a simple second-order law,
\bq R(x) \equiv k_R n p \eq
The electrons are assumed to be at quasiequilibrium throughout the semiconductor, so that
\bq n(x) = \nb e^{\phi} \eq
The hole distribution in the bulk will be computed explicitly below. In the depletion region, the transport equations can be solved to relate the hole concentration profile $p(x)$ to the concentration at the edge of the depletion region $p_w$,
\bq p(x) = p_w e^{-\phi} + p^\dagger(x) \eq
where
\bq p^\dagger(x) = e^{-\phi(x)} \int_{-w}^x \theta(x') e^{\phi(x')} \, dx' \eq
\bq \theta(x) = - F_p + \int_x^0 G(x') - R(x') \, dx' \eq
Note that $p_\dagger$ represents the deviation from quasiequilibrium, as setting $p^\dagger = 0$ is equivalent to assuming that the holes are at quasiequilibrium throughout the depletion region. An analysis of this integral is given in the SI, and shows that we may use the quasiequilibrium approximation,
\bq p(x) \approx p_w e^{-\phi} , \label{pxqeq} \eq
with error of order $\lambda/\delta$; this approximation therefore holds when the Debye length is much smaller than the diffusion length. In the SI it is shown that, even when this assumption is relaxed, the current takes the same form in that it is linear in $J_G$ and $p_w$. Berz has also given an analysis of the validity of the quasiequilibrium assumption \cite{berz79} using a different method. It is important to note that this assumption changes the form of the depletion region recombination current relative to other classical treatments.

The hole continuity equation is
\bq \frac{1}{q} \frac{d J_p}{dx} = G(x) - R(x) \eq
Integrating across the depletion region,
\bq J_p = J_\text{bulk} + J_{\Phi,\text{dep}} - J_{R,\text{dep}} \label{jpgar} \eq
where $J_p=J(0)$ is the total hole current passing through the surface, $J_\text{bulk} = J(-w)$ is the hole current from the bulk to the depletion region,
\bq J_{\Phi,\text{dep}} \equiv q \Phi_\text{dep} = q \int_{-w}^0 G \, dx = q \Phi ( 1-e^{-\alpha w} ) \eq
is the current generated by illumination in the depletion region, and
\bq J_{R,\text{dep}} \equiv q \int_{-w}^0 R \, dx = q k_R w \nb p_w \eq
is the depletion recombination current.

$J_\text{bulk}$ is obtained by assuming that there is no field in the bulk and that the electron concentration remains unperturbed from its equilibrium value $\nb$. Therefore in the bulk the hole continuity equation is
\bq - \frac{d^2 p}{dx^2} = \frac{\Phi \alpha e^{\alpha x}}{D_p} - \frac{p-\pb}{\delta^2} \eq
Solving this with boundary conditions $p(-\infty) = \pb$ and $p(-w)=p_w$ gives the solution for $J_\text{bulk} = q(dp/dx)(w)$,
\bq J_\text{bulk} = J_{\Phi,\text{bulk}} + J_{D,\text{bulk}} . \eq
The current due to generation in the bulk is
\bq J_{\Phi,\text{bulk}} = q \eta_\text{bulk} \Phi_\text{bulk} , \eq
where
\bq \Phi_\text{bulk} = \Phi e^{-\alpha w} \qquad \eta_\text{bulk} = \frac{\alpha \delta}{1+\alpha \delta} . \eq
Here $\eta_\text{bulk}$ is the fraction of charges generated in the bulk that reach the edge of the depletion region before recombining. The diffusion current term is
\bq J_{D,\text{bulk}} = q D_p \frac{\pb-p_w}{\delta} \eq
which is the current due to diffusion across one diffusion length just inside the bulk region, as depicted in Figure \ref{holediag}. The concentration profile shown here is a schematic one; in reality, the hole concentration varies throughout the bulk and in general has a nonlinear profile that is dependent on the magnitude of generation and recombination, but the diffusional current is mathematically equivalent to the simple conceptual illustration in the figure. Note that the original G\"artner model assumes $p_w=0$.

Substituting these results into \er{jpgar},
\bq \begin{aligned} J_p &= -q \Phi \left( 1 - \frac{e^{-\alpha w}}{1+\alpha \delta} \right) + q D_p \frac{\pb-p_w}{\delta} - q k_R w \nb p_w \\ &= J_G - q \left( \frac{D_p}{\delta} + k_R w \nb \right) p_w \end{aligned} \eq
The result is the original G\"artner current $J_G$ minus an extra term proportional to $p_w$ that describes additional recombination losses due to the hole transport limitation. The first term, $qD_p p_w/\delta$, is the amount of current fed back into the bulk, where the holes recombine, and the second term, $q k_R w \nb p_w$, is the amount of current lost due to recombination in the depletion region.

In the original G\"artner model the diffusion current $q D_p \pb/\delta$ is typically not the major contribution to the total current because of the relatively small value of $\pb$; however, if the kinetics are slow and there is a large buildup of holes in the depletion region, there may be enough of a back current that $p_w$ exceeds $\pb$, leading to a net negative diffusion current. If $p_w$ becomes large enough, the diffusion current may eventually eclipse the generation current; when this occurs we say that the current becomes limited by the hole transport. Depending on the relative values of $D_p/\delta$ and $k_R w \nb$, a large $p_w$ may also limit the hole current through recombination in the depletion region.

To couple this to the boundary conditions, we need to be able to relate $p_w$ to $p_s$. Using the quasiequilibrium assumption for the hole concentration profile \er{pxqeq}, we have $p_s = p_w e^{\vsc}$, and we may finally write
\bq J_p = J_G - \bar{J}_R e^{\bar{V}_\text{sc} - \vsc} p_s / \pb_s \label{jp} \eq
where
\bq \bar{J}_R = q \left( \frac{D_p}{\delta} + k_R w \nb \right) \pb_s e^{-\bar{V}_\text{sc}}\eq
is the depletion back-current and recombination current at equilibrium. (Note that due to the appearance of $w$, this quantity is not exactly constant, but can be treated as such for practical purposes.)

By using the two approximations Eqs.\ \eqref{nsqeq} and \eqref{jp} for the surface electron and hole concentrations, the numerical simulation can be dispensed with and the semiconductor transport, generation, and recombination processes can be described analytically. Because this generalization of the G\"artner model involves additional recombination, either in the form of holes passing from the depletion region back into the bulk and recombining there or recombining in the depletion region itself, we refer to the sum of these two effects as ``depletion recombination.'' 

\subsection{Interfacial electron transfer: surface states, catalyst, and solution}

In some previous models it has been assumed that surface recombination follows simple second order kinetics so that the recombination current is $J_r = k_r n_s p_s$ \cite{peter12}. In others the filling of the surface states is calculated, but the reverse reactions are not considered \cite{ponomarev95}. This is problematic because without the reverse reactions, the model cannot correctly describe quasiequilibrium, which can occur when semiconductor-surface state transfer is rapid.

The density of states of the surface states, catalyst, and solution all play a role in determining the form of the current. It has been suggested that an exponential density of surface trap states is required to quantitatively model the photocurrent response of TiO$_2$ \cite{wilson77}, and the solution density of states can be modeled with the Marcus-Gerischer model \cite{vanmaekelber97}. These effects can be included in this model, but doing so would only obfuscate the results. For clarity we use a single-energy surface state model (as in the Shockley-Read-Hall model) and broad-DOS catalyst and solution models.

\subsubsection{Electron transfer model}

The model for interfacial electron transfer is based on a generalization of simple second-order reaction kinetics. For transfer between subsystems 1 and 2, we write $d_i(\e)$ for an electron donor species and $a_i(\e)$ for an electron acceptor species in subsystem $i$ and electron energy $\e$. At each value of $\e$, the basic reaction is
\bq a_1(\e) + d_2(\e) \longleftrightarrow d_1(\e) + a_2(\e) \eq
with the reaction proceeding to the right representing positive current from subsystem 1 to 2. The current, proportional to the total reaction rate, is computed by integrating the rate densities over the electron energy $\e$,
\bq J_{1,2} = q \int k_{1,2}(\e) \left[ a_1(\e) d_2(\e) - d_1(\e) a_2(\e) \right] \, d\e \label{jint1} \eq
The donor and acceptor distributions can be written as the product of an electronic density of states (DOS) function $g_i(\e)$ and an occupancy probability (Fermi-Dirac) function $f_i(\e)$, where $f_i(\e) = 1/(1+e^{\e-E_i})$;
\bq d_i(\e) = g_i(\e) f_i(\e) \quad a_i(\e) = g_i(\e) [ 1 - f_i(\e) ] \eq
Substitution gives for the current integral
\bq J_{1,2} = q \int k_{1,2}(\e) g_1(\e) g_2(\e) \left[ f_1(\e) - f_2(\e) \right] \, d\e \label{jint2} \eq
The DOS function used for the semiconductor and catalyst are constants; for the surface states, an impulse function [$g_\text{ss}(\e) = \nss \delta(\e-\esso)$]; and for the solution, the large-$\lambda$ limit of the Marcus-Gerischer DOS,\cite{Marcus64,gerischer61}
\bq d_\text{sol}(\e) = c e^{-(\e-\esol)/2} \quad a_\text{sol}(\e) = c e^{-(\esol-\e)/2} \eq
The interfacial current expressions given below are all derived from this model by evaluating the current integrals Eqs.\ \eqref{jint1} and \eqref{jint2}; see the SI for more details.

We will write $s$ and $s^+$ for the neutral and oxidized surface state concentrations, respectively, where
\bq s = \nss f(\ess-\esso) \qquad s^+ = \nss-s \eq
We will also write $c_\text{ss}$ and $c_\text{ss}^+$ for the occupation of catalyst sites at energy $\esso$,
\bq c_\text{ss} = f(\ecat-\esso) \qquad c_\text{ss}^+ = 1-c_\text{ss} \eq
These quantities occur when discussing surface state-catalyst transfer. We also define $\Delta \vh \equiv \vh-\vho$ as the deviation of the Helmholtz potential from its equilibrium value.

\subsubsection{Interfacial currents}

\begin{figure}\begin{center}\includegraphics[scale=0.85]{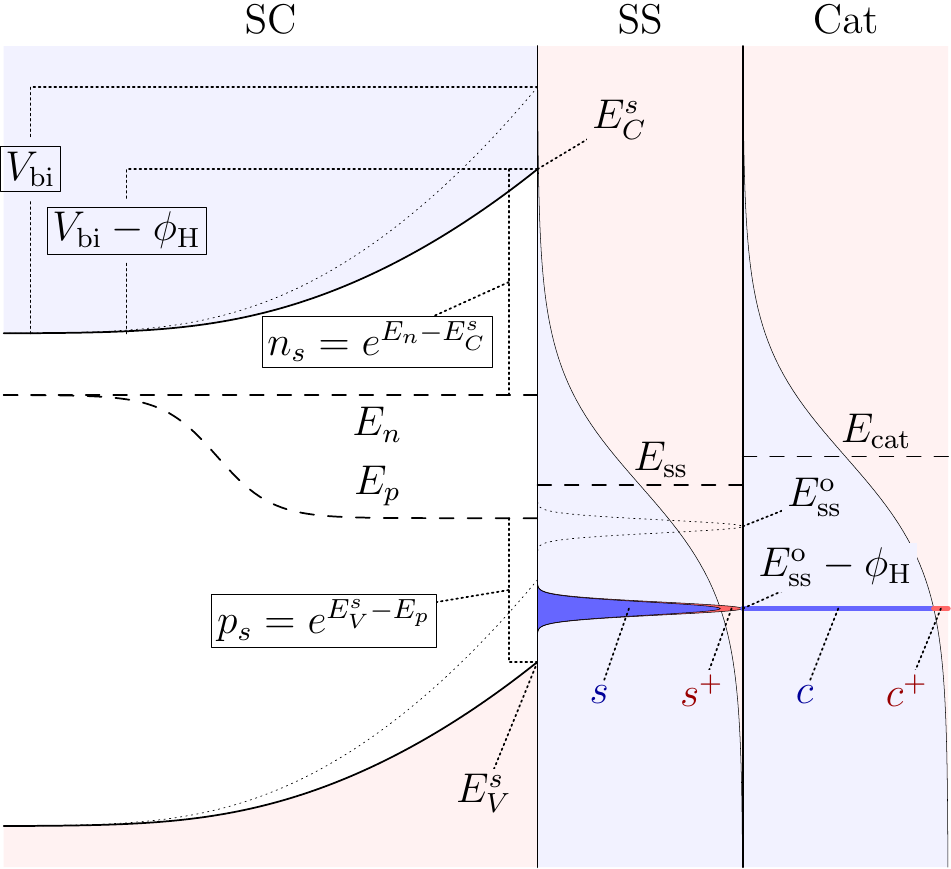} \caption{Energy diagram of semiconductor and catalyst, showing the shifting of the semiconductor bands and the surface state energy due to the Helmholtz potential. Transfer between surface states and catalyst can only occur at the shifted surface state energy $\bar{E}_\text{ss}^o - \phi_\text{H}$.} \label{sccurdiag}\end{center}\end{figure}

The reaction of semiconductor holes and electrons with surface states follows simple second-order kinetics,
\bq J_\vbss = \jo_\vbss \left( \frac{p_s s}{\pb_s \bar{s}} - \frac{s^+}{\bar{s}^+} \right) \eq
\bq J_\cbss = \jo_\cbss \left( \frac{s}{\bar{s}} - \frac{n_s s^+}{\nb_s \bar{s}^+} \right) \eq

Reaction of surface states with catalyst and solution must take into account the fact that the surface states are at an electrostatic potential $-\vh$ with respect to the catalyst and solution (see Sec \ref{epot}); this shift puts the surface state energy at $\esso=\besso-\vh$. The current then depends on the surface state concentration and the catalyst occupancy at energy $\esso$;
\bq J_\sscat = \frac{\jo_\sscat}{\bar{s}^+ \bar{c} } \left( s^+ c_\text{ss} - s c_\text{ss}^+ \right) \eq
The surface state-solution current is
\bq J_\sssol = \jo_\sssol \left( \frac{s^+}{\bar{s}^+} e^{-\Delta \vh/2} - \frac{s}{\bar{s}} e^{\Delta \vh/2} \right) \eq

The expressions we used previously \cite{mills14} to model the current between semiconductor and ion-permeable catalyst with broad DOS (``adaptive'' catalyst) have to be modified to account for the electrostatic potential drop between the semiconductor surface and the catalyst, $\vh$:
\bq J_\vbcat = \jo_\vbcat \left( \frac{p_s}{\pb_s} - e^{-\Delta \vh-\ecat} \right) \eq
\bq J_\cbcat = \jo_\cbcat \left( e^{\Delta \vh+\ecat} - \frac{n_s}{\nb_s} \right) \eq
The catalyst-solution current
\bq J_\catsol = \jo_\catsol \left( e^{-\ecat/2} - e^{\ecat/2} \right) \eq
does not depend on $\vh$ because the catalyst and solution always remain at the same potential (due to the screening of catalyst charge by ions in the permeable catalyst).

\subsection{Surface state occupancy, recombination, and transfer} \label{ssmod}

Our surface state model is closely related to the Shockley-Read-Hall (SRH) recombination model as we have used a single energy level DOS for the surface states. However, because we allow for transfer between the surface states, catalyst, and solution, the steady-state occupancy of the surface states is different from that predicted by the SRH model. Some of the positive charge injected into the surface states does not participate in recombination but instead proceeds further to oxidize the catalyst and/or solution; this effect has been discussed before by van Maekelbergh \cite{vanmaekelber97} and is called \textit{surface state-mediated transfer}. The results of this section are not required for solution of the model equations, but illuminate the relationship between our model and the SRH model and quantify the effect of surface state-mediated transfer.

The occupancy of the surface states is determined by applying current equality through them;
\bq J_\vbss + J_\cbss = J_\sscat + J_\sssol \label{ssjeq} \eq
Solving for $s^+$ and calculating the currents permits one to write them in the form
\bq J_\vbss = J_r^\text{ss} + J_\vbcat^\text{ss} + J_\vbsol^\text{ss} \eq
\bq J_\cbss = -J_r^\text{ss} + J_\cbcat^\text{ss} + J_\cbsol^\text{ss} \eq
Here, $J_r^\text{ss}$ represents the surface state recombination current, and the others represent surface state-mediated transfer, i.e. the current passed through the surface states from the semiconductor into the catalyst and solution. These currents are
\bq J_r^\text{ss} = u_\text{ss} \jo_\vbss \jo_\cbss \left( \frac{p_s n_s}{\pb_s \nb_s} - 1 \right) \label{jrss} \eq
\begin{multline} J_\text{sc,cat}^\text{ss} = u_\text{ss} \jo_\sscat \left[ \jo_\vbss \left( \frac{p_s c}{\pb_s \bar{c}} - \frac{c^+}{\bar{c}^+} \right) \right. \\ \left. + \jo_\cbss \left( \frac{c}{\bar{c}} - \frac{ n_s c^+}{\nb_s \bar{c}^+} \right) \right] \label{ssjcat} \end{multline}
\begin{multline} J_\text{sc,sol}^\text{ss} = u_\text{ss} \jo_\sssol \left[ \jo_\vbss \left( \frac{p_s}{\pb_s} e^{\Delta \vh/2} - e^{-\Delta \vh/2} \right) \right. \\ \left. + \jo_\cbss \left( e^{\Delta \vh/2} - \frac{n_s}{\nb_s} e^{-\Delta \vh/2} \right) \right] \label{ssjsol} \end{multline}
where
\begin{multline} u_\text{ss} = (1+ e^{-\esso}) \left[ \jo_\vbss \left( \frac{p_s}{\pb_s} + e^{-\esso} \right) \right. \\ \left. + \jo_\cbss \left(1 + \frac{n_s}{\nb_s}e^{-\esso} \right) + \jo_\sscat (1+e^{-\esso}) \right. \\ \left. + \jo_\sssol \left( e^{-\Delta \vh/2} + e^{\Delta \vh/2 - \esso} \right) \right]^{-1} \end{multline}
\er{jrss} is analogous to the main result of the SRH model, but is modified by the factor $u_\text{ss}$, which decreases when charge is transferred through the surface states instead of recombining. This factor essentially partitions the current into the surface states between recombination and further transfer out of the states. It is important to note that the presence of the catalyst can decrease the recombination current relative to a system without catalyst by moving charge out of the surface states and into the catalyst or solution, raising the surface state potential $\ess$ hence reducing the states, leaving fewer holes in the surface states to recombine with electrons from the conduction band. The transfer currents Eqs.\ \eqref{ssjcat} and \eqref{ssjsol} take the form of second-order rate expressions for direct transfer between semiconductor, catalyst, and solution, with more complicated ``exchange currents'' that depend on the applied bias through the factor $u_\text{ss}$.

We note also that the same analysis can be applied to the catalyst, which can function both as a recombination center and as an intermediary by which charge can be passed from the semiconductor to the solution, in the same way that surface states can. However, because of the different DOS of the catalyst, the analysis is more involved, but the basic mechanisms and conclusions are the same.

\subsection{Solution of the model equations}

There are four variables in the model: $p_s$, $\ess$, $\ecat$, and $\vh$. The four equations required to obtain the solution are the electroneutrality condition \er{potqeq} and the subsystem current equality conditions
\bq J_\vbss + J_\cbss = J_\sscat + J_\sssol \eq
\bq J_\vbcat + J_\cbcat + J_\sscat = J_\catsol \eq
\bq J_p = J_\vbss + J_\vbcat \eq
where $J_p$ is given by \er{jp}. This last equation can be solved to get the surface hole concentration,
\bq \frac{p_s}{\pb_s} = \frac{J_G + \jo_\vbss s^+/\bar{s}^+ + \jo_\vbcat e^{-\Delta \vh - \ecat} } { J_\text{dr} e^{-\Delta \vsc} + \jo_\vbss s/\bar{s} + \jo_\vbcat} , \eq
leaving the remaining three to be solved numerically. This system presents numerical challenges in its solution because it involves combinations of variables that may vary by many orders of magnitude. The method we adopted is to numerically approximate the equilibrium Helmholtz potential and open-circuit voltage, begin the numerical solution at this applied bias with all potentials set to zero, then scan the applied bias in small increments away from there, using the previous solution as an initial guess for the next step. More details and a Mathematica implementation of this algorithm are included in the SI.

\section{Results/Discussion}

Due to the large number of parameters - in particular, seven exchange currents and the surface state parameters $\nss$ and $\besso$ - the behavior of the combined system is very complicated. In this section we separate the effects of the surface states and the catalyst in order to clarify the interactions between the two. We first show that systems with only either catalyst or surface states are closely related to our previous adaptive, metallic and molecular models \cite{mills14}. We then explore catalyst-surface state interactions by investigating the limits of high and low semiconductor- and catalyst-surface state exchange currents, which correspond to series and parallel effects of the catalyst. In any given physical system both effects will operate simultaneously, but this separation allows us to demonstrate the different mechanisms by which catalyst overlayers can increase both the attainable current density and photovoltage.

In this section we will give some analytical results in addition to plots and qualitative descriptions of the simulation data. These equations and approximations demonstrate how our analytical model can be used to derive quantitative predictions of certain limits of the model, but they are not necessary to understand the simulation results and the main conclusions of this work. The SI contains more details on the derivation of these results.

\subsection{Non-ideal photodiodes}

\begin{figure}\begin{center}\includegraphics[scale=0.8]{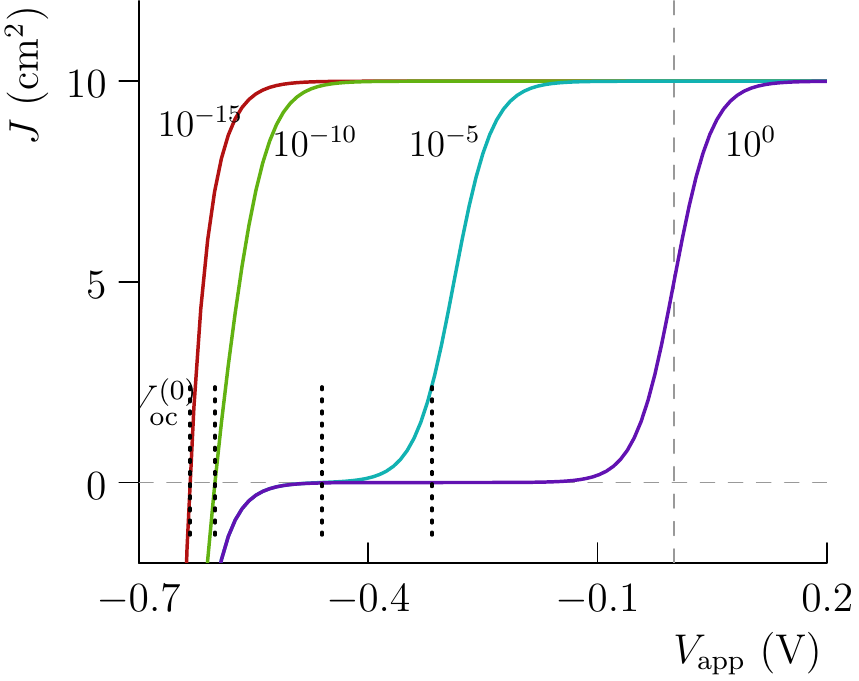} \caption{$J(V)$ response of a "non-ideal" photodiode, the limiting current obtainable from a fast catalyst, for different values of $\jo_R/\jo_\text{vb,sol}$ (marked).} \label{idiode}\end{center}\end{figure}

The addition of the hole transport limitation and depletion region recombination decreases the maximum current obtainable from the semiconductor under otherwise ideal conditions (i.e.\ fast OER kinetics at the surface states and/or catalyst). Before discussing the roles of surface states and catalyst, we analyze the deviation of the semiconductor response from its ideal behavior due to slow hole transfer out of the semiconductor.

The ideal photodiode equation is a simple model that describes the $J(V)$ behavior obtained from an ideal system (fast hole transfer from the semiconductor and fast OER kinetics),
\bq J_\text{id} = J_G - \jo_\text{cb,sol} e^{-V} \eq
which results from assuming a constant forward hole current $J_G$ (ignoring hole back-current), quasiequilibrium of electrons in the semiconductor, and quasiequilibrium of the surface states and catalyst with the solution, where $\jo_\text{cb,sol} = \jo_\cbss + \jo_\cbcat$ is the effective exchange current. It provides a simple means of estimating $\voc$,
\bq \voc^{(0)} \approx -\ln\left(\frac{J_G}{\jo_\text{cb,sol}} \right) \eq
However, in the presence of depletion layer recombination, the hole current must be modified according to the generalized G\"artner model, \er{jp}. This gives the "non-ideal" photodiode equation,
\bq J_\text{n-id} = \frac{J_G}{1+(\jo_R/\jo_\text{vb,sol}) e^{-V}} - \jo_\text{cb,sol} e^{-V} \eq
which is shown in Figure \ref{idiode} for various values of $\jo_R/\jo_\vbsol$. This is the highest current obtainable from the semiconductor in the presence of depletion recombination. Including this effect leads to a shift of $\voc \approx \voc^{(0)} + \voc^{(1)}$, where
\begin{multline} \voc^{(1)} = -\ln \left[\left(\sqrt{\frac14 + \frac{J_G \jo_R}{\jo_\text{vb,sol} \jo_\text{cb,sol}}}-\frac12\right)\right. \\ \left.\cdot\left(\frac{\jo_\text{vb,sol} \jo_\text{cb,sol}}{J_G \jo_R}\right)\right] \label{voc1} \end{multline}
Note that the decrease in current is not due to increased electron transfer to the solution, and does not cause any change to the band bending; it is due only to increased recombination internal to the semiconductor caused by a buildup of holes in the depletion region.

\subsection{Basic transfer models}

\begin{figure}\begin{center}\includegraphics[scale=0.7]{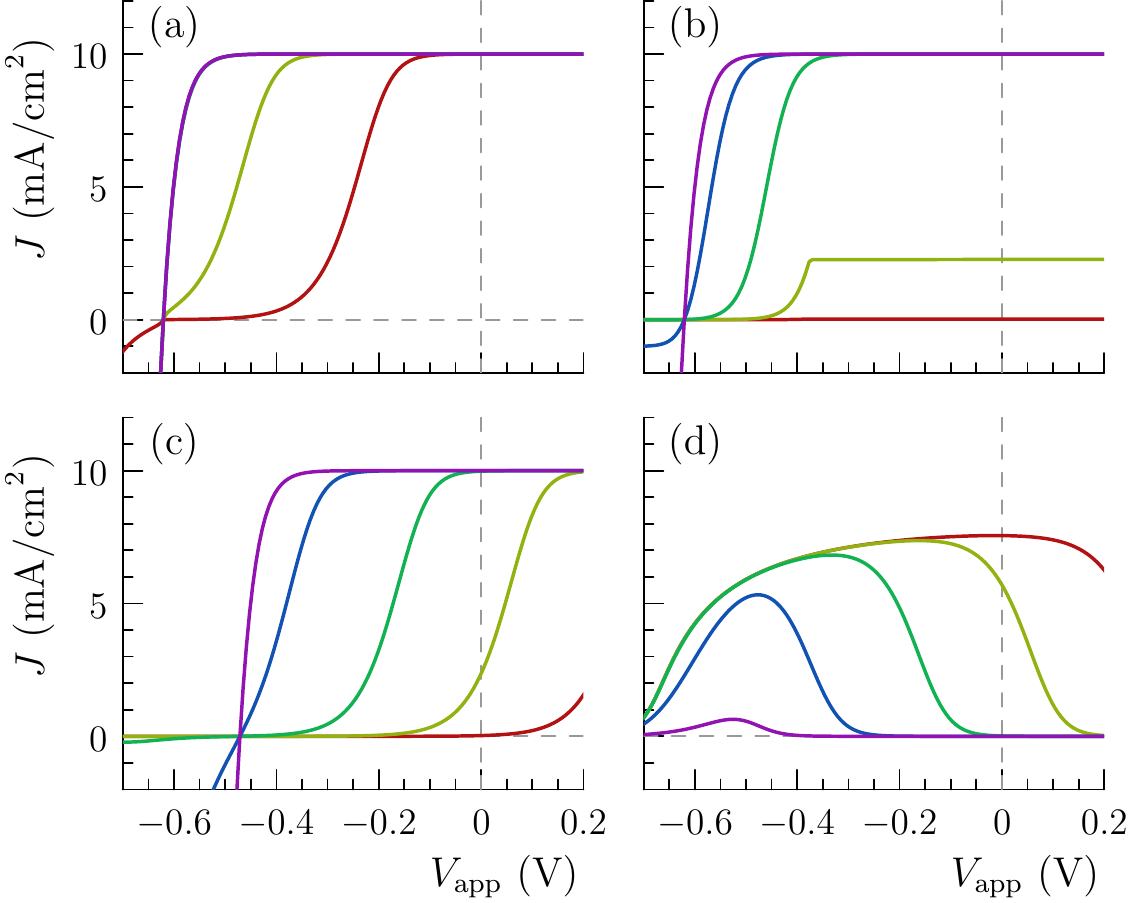} \caption{(a) Permeable catalyst, no surface states --- "adaptive" model (Sec \ref{adapcat}); (b) Screened surface states, no catalyst --- "molecular" model (Sec \ref{molcat}); (c) Unscreened surface states, no catalyst -- "metallic" model with shifted barrier height (Sec \ref{metcat}); (d) Surface recombination current [\er{jrss}] for the system in (c).} \label{jfig1}\end{center}\end{figure}

We begin by investigating the transfer mechanisms of systems comprised of a semiconductor and either catalyst or surface states alone. We show that these correspond to the adaptive, metallic, and molecular catalyst models we defined and analyzed previously in our simulation work. Figure \ref{jfig1} shows the currents obtained in these models. Each color in each graph corresponds to the same value of exchange current with the solution, $\jocs$ or $\joss$; in this section these exchange currents vary from $10^{-6}$ to $10^{2}\,\text{mA}\,\text{cm}^{-2}$ in steps of $10^2$ (other parameters are found in the SI). Plots of the relevant potentials and discussion of the behavior of each model are presented below. Above each potential plot is a band diagram of the system showing the potentials at a selected bias, indicated on the figures.

\subsubsection{Adaptive catalysts --- no surface states}\label{adapcat}

\begin{figure}\begin{center}\includegraphics[scale=1]{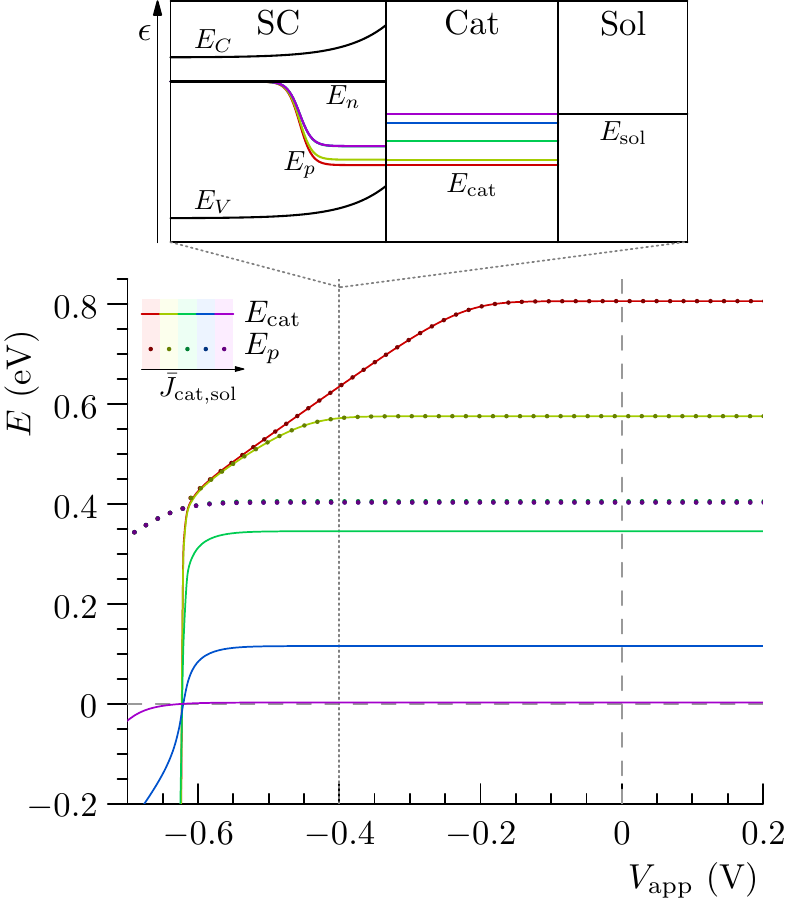} \caption{$\ecat$ and $E_p$ for the adaptive catalyst model [see Fig \ref{jfig1}(a), Sec \ref{adapcat}].} \label{e1fig}\end{center}\end{figure}

First we consider a system with a permeable catalyst layer and neglect surface states, as in our previously defined "adaptive" catalyst model. The results obtained with the analytical model are the same as those obtained by our previous simulations. The catalyst potential shifts in order to accomodate slower catalysts (smaller $\jocs$), until depletion recombination sets in and decreases the hole current, as can be seen in Fig \ref{e1fig} which shows $\ecat$ and $E_p$. With high values of $\jocs$, $\ecat$ remains close to $\esol$. As $\jocs$ decreases, the catalyst potential shifts to increase the reaction rate and compensate for the slower catalyst. At low enough values of $\jocs$, the current is limited by the rate of hole transfer out of the semiconductor, i.e.\ $\ecat$ approaches $E_p$. When the surface hole density required for passing current $J_G$ becomes high enough, it becomes limited by depletion recombination, leading to a much slower increase of $E_p$, hence $\ecat$ and therefore the total current.

These key potentials can be quantified. When $\vap$ is high enough to obtain $J=J_G$, $\ecat$ levels off to a constant value,
\bq \ecat[J_G] \approx 2 \ln \left( \frac{J_G}{\jo_\catsol} \right) \label{ecatjg} \eq
However, the bias at which this potential is reached is limited by depletion recombination. When depletion recombination occurs, the maximum $\ecat$ at bias $V$ is
\bq \ecat[\text{max}] \approx V + \ln \left( \frac{J_G}{\jo_R} \right) \eq
This is the diagonal line in Fig \ref{e1fig}. When depletion recombination occurs, the current will not reach $J_G$ until the bias reaches
\bq \vap[J_G] = \ln \left( \frac{J_G^3}{\jo_R\jo_\catsol^2} \right) \eq
This bias can be seen in Figs \ref{jfig1}(a) and \ref{e1fig}. This equation can be used to determine if a system will be affected by surface recombination, if $\vap[J_G] > \voc$, where $\voc$ is given by \er{voc1}. Note the square in the deminator; this explains why $J$ shifts at a more rapid rate than might be expected from the value of $\jo_\catsol$.

At high values of $\jocs$, the hole concentration is determined by whatever is necessary to pass current $J_G$ from the valence band to the catalyst,
\bq E_p[\text{min}] = \ln \left( \frac{J_G}{\jo_\vbcat} \right) \eq
However, when the catalyst is slow, a greater hole concentration is necessary to reach $J_G$. In this case $E_p$ tends to $\ecat[J_G]$; in fact, 
\bq E_p \approx \ln \left[ \frac{J_G}{\jo_\vbcat} + \left( \frac{J_G}{\jocs} \right)^2 \right], \eq
showing the interpolation between the minimum value and $\ecat$.

A crucial aspect of adaptive catalysts that allows them to respond so effectively to a low $\jocs$ is that $\ecat$ can swing very quickly from negative to positive values near $\voc$, a property not shared by the other catalyst models. Indeed, the value of $\ecat$ as a function of applied bias (above $\voc$) without depletion recombination is
\bq \ecat[\vap] \approx 2 \ln \left( \frac{J_G - \jo_\cbcat e^{-\vap}}{\jocs} \right) \label{ecatv} \eq
This function swings very rapidly from negative values to $\ecat[J_G]$ as $\vap$ passes through $\voc$, as reflected in Fig \ref{e1fig}.

\subsubsection{Screened surface state transfer --- molecular catalyst}\label{molcat}

\begin{figure}\begin{center}\includegraphics[scale=1]{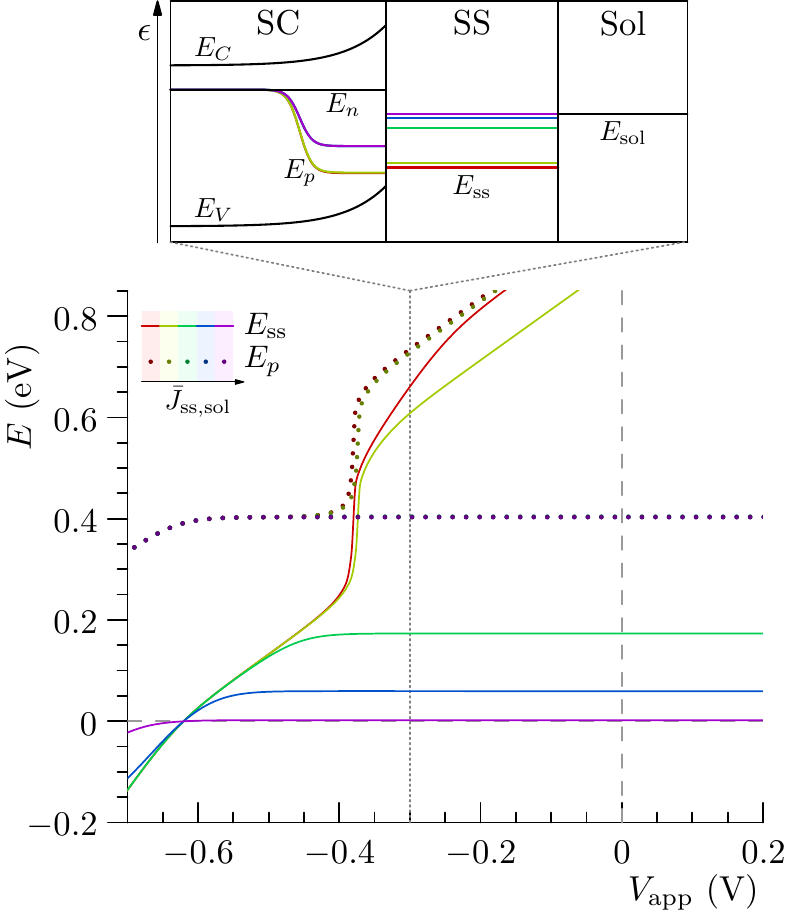} \caption{$\ess$ and $E_p$ for the screened surface state "molecular" model [see Fig \ref{jfig1}(b), Sec \ref{molcat}].} \label{e2fig}\end{center}\end{figure}

Next we consider transfer from semiconductor surface states directly to solution without a catalyst layer. Such "catalytic" surface states correspond to e.g.\ high-oxidation state metal centers in iron oxide, or a surface-attached molecular redox species. The surface states are said to be screened when they do not create a substantial Helmholtz potential. This could occur because they protrude into the electrolyte, as in surface-attached molecules, or because the surface state charge is negligible in comparison to the charge in the Helmholtz layer. The latter case occurs, for example, when the redox potential of the surface states $\besso$ is more negative than $\esol$; at equilibrium, the states remain mostly uncharged.

A system with surface states situated at $\esso = -.25\text{ eV}$ below $\esol$ is depicted in Figs \ref{jfig1}(b) and \ref{e2fig}. Immediately we can see that, although there still is no Helmholtz potential, the molecular catalyst has slightly poorer performance than adaptive catalysts at a given value of $\joss$. At low enough values of $\joss$, the catalyst saturates ($s^+ \approx \nss$), and is unable to provide enough charge to the solution for the current to reach $J_G$.

These effects are a direct consequence of the localized density of states of a molecular catalyst. In the adaptive model, where there is a broad DOS, the electrons and holes in the semiconductor can essentially transfer charge independently (i.e.\ have first-order kinetics), whereas with a localized catalyst DOS, they both must pass through the same energy state and are therefore more tightly coupled (have second-order kinetics). This means that the electrons can more readily reduce the charge in the surface states (surface recombination), leading to a lower solution current.

With screened surface states - ``molecular catalysts'' - as the active OER sites, the potential $\ess$ (above $\voc$) in the absence of depletion recombination is
\bq \ess(V) \approx \ln \left( \frac{J_G}{\joss + \jo_\cbss e^{-\vap}} \right) \label{essvmol} \eq
Contrast this with \er{ecatv}. Here, the term containing $e^{-\vap}$ in the denominator limits the potential to the line $\vap+\voc$, as seen in Fig \ref{e2fig}. This is a reflection of the second-order kinetics between electrons and surface states, absent in the adaptive model. Note also that the maximum value here, $\ess[\text{max}] = \ln(J_G/\jo_\sssol)$, is less than the corresponding value \er{ecatjg}, because of the factor of $2$ present there - this is again a consequence of the different densities of states, as the adaptive catalyst follows Butler-Volmer-type kinetics with transfer coefficient $1/2$, whereas the molecular catalyst follows the form typical of semiconductors or redox couples with transfer coefficient $1$.

In this system, $\ess$ increases too slowly with bias to reach a high enough hole concentration to cause depletion recombination before the states become completely filled, seen as the abrupt increase in $\ess$ and $E_p$ in Fig \ref{e2fig} near $-0.4\text{ V}$. At this point, the hole concentration increases rapidly in an attempt to produce as much current as possible and reaches the limit determined by depletion recombination [$E_p < \vap + \ln(J_G/\jo_R)$], but since there are no more available surface states, this has no effect on the current. Depletion recombination could be observed in such a system with a higher value of $\jo_R$, a lower value of $\jo_\vbss$, a lower value of $\voc$, etc.

\subsubsection{Unscreened surface state transfer --- metallic catalyst}\label{metcat}

\begin{figure}\begin{center} \includegraphics{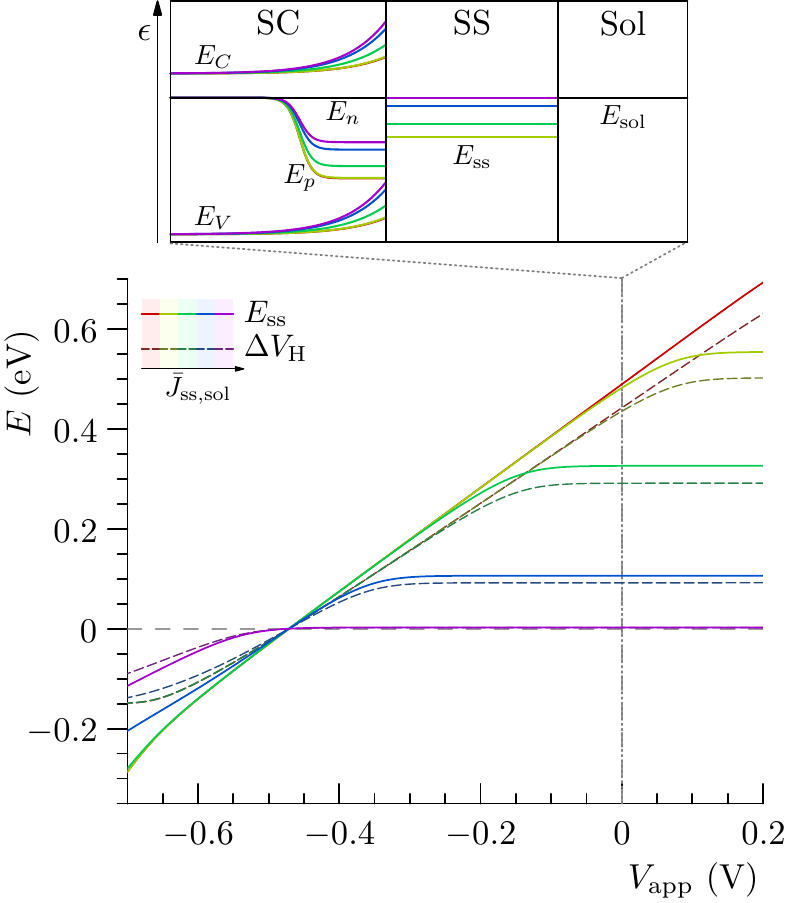} \caption{$\ess$, $\vh$, and $E_p$ for the unscreened surface state "metallic" model [see Fig \ref{jfig1}(c) and (d), Sec \ref{metcat}].} \label{e3fig}\end{center}\end{figure}

In many systems with surface states, particularly those with higher redox potentials ($\esso > \esol$), the states will be partially filled almost always, and when these cannot be screened by ions (as with surface species embedded in the semiconductor rather than surface-attached species), they will produce a substantial Helmholtz potential. This potential decreases the amount of band bending in the semiconductor, leading to more electron and less hole current, and hence generally worse performance.

The presence of the surface states and the Helmholtz potential causes the system to act in many ways analogously to the ``metallic'' catalysts we have previously defined. The physical analogy is that there is a large density of states whose chemical potential is essentially controlled by the electrostatic potential $\vh$. This is the same physical situation as found in metallic catalysts, which also contain a large quantity of unscreened charge states and hence can only be affected by a change in the electrostatic potential drop between the catalyst and the solution.

The equilibrium Helmholtz potential $\vho$ was discussed in Sec \ref{epot}, and is always greater than $\besso$, the amount depending on $\nss$. This potential is essentially lost to the system; the system behaves as though the equilibrium barrier height is smaller by $\vho$. This leads to a shift in $\voc$ and the $J(V)$ response, as seen in Fig \ref{jfig1}(c). Hence, these systems behave in the same way as those with a metallic catalyst whose Fermi level differs from the solution potential by the amount $\vho$.

The decreased band bending in systems with metallic catalysts leads to an earlier onset of electron current as $\joss$ decreases, because of the greater electrostatic potential required to drive the reaction. Before the current reaches $J_G$, we have
\bq \vh \approx \vap +\ln \left( \frac{J_G}{\jo_\cbss} \right) = \vap + \voc \eq
similar to the way that the molecular catalyst potential $\ess$ is limited by $\vap+\voc$ [see \er{essvmol}], albeit for a different reason. In this regime,
\bq \begin{aligned} \ess & \approx \esso - \vh + \ln\left( \frac{\nss}{\ch \vh} - 1 \right) \\ & \approx V + \voc - \vho + \ln\left[ \esso (V + \voc) \right] \end{aligned} \eq
This small deviation from $V+\voc-\vho$ is visible in Fig \ref{e3fig}, and represents the minor deviation from the exact metallic model. It is a result of the interaction between charge neutrality and catalyst kinetics.

In other respects it behaves like the metallic model, except that the transfer coefficient in this case is 1 instead of $1/2$ due to the localized DOS. The potential at $J=J_G$ is
\bq \ess = \esso + \ln\left(\frac{\joss e^{-\Delta V_H}(1+e^{\esso})}{J_G}-1\right) \eq
and the bias required to reach this current is
\bq V = \bvh+\ln\left[\frac{\joss\jo_\cbss}{J_G^2}\left(\frac{s^+}{\bar{s}^+}\right)\right] \eq

Figure \ref{jfig1}(d) shows the surface recombination current \er{jrss}. The onset of surface recombination shifts with $\joss$ as a result of the shift in $\vsc$ and the consequent shift in onset of electron current. Note that there is substantial recombination current even when the total current is near zero; in this regime, the electron current is balanced by the hole current, so all of it results in surface recombination.

\subsection{Catalyst overlayer effects}

\begin{figure}\begin{center}\includegraphics[scale=0.7]{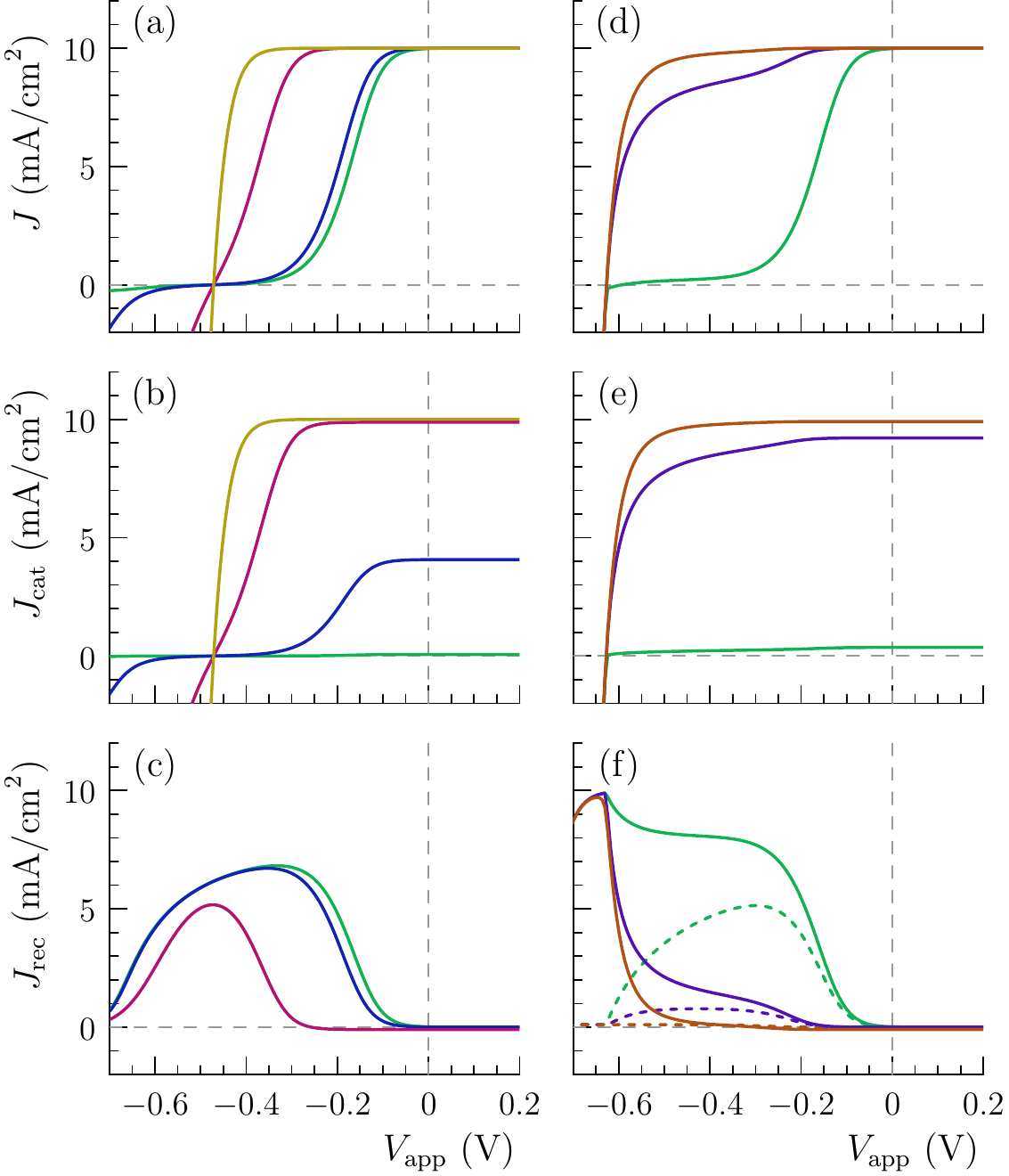} \caption{(a), (b), (c) -- total current $J$, catalyst-solution current $J_\text{cat}$, and recombination current $J_\text{rec}$ for the series catalyst model (Sec \ref{sercat}), with $\jocs=10^{-4}, 10^{-2}, 10^0, 10^{2}$; (d), (e), (f) -- $J$, $J_\text{cat}$, $J_\text{rec}$ for the parallel catalyst model (Sec \ref{parcat}), with $\jocs=10^{-4}, 10^{-2}, 10^0$. In (f), dotted lines show surface state recombination only.} \label{jfig2}\end{center}\end{figure}

We now study the interaction between the surface states and catalyst overlayer by selecting the system represented by the green curve, $\joss=10^{-2}\,\text{mA}\,\text{cm}^{-2}$, from the unscreened surface state (metallic) model above [Fig \ref{jfig1}(c), Sec \ref{metcat}], and adding a catalyst overlayer to it. The green curves in the figures below correspond to this ``parent'' curve, i.e.\ with negligible catalyst contribution to the total current. We discuss three relevant examples in order to demonstrate the interaction effects: a model in which the catalyst operates primarily in series with the surface states, one in which it operates primarily in parallel, and one in which it does not transfer to the solution (non-catalytic overlayer). Figure \ref{jfig2} presents the total, catalyst-solution, and recombination currents for the series and parallel models. As in the previous section, band diagrams at selected biases are shown above the potential diagrams.

\subsubsection{Series effect --- surface-state mediated transfer}\label{sercat}

\begin{figure}\begin{center}\includegraphics[scale=1]{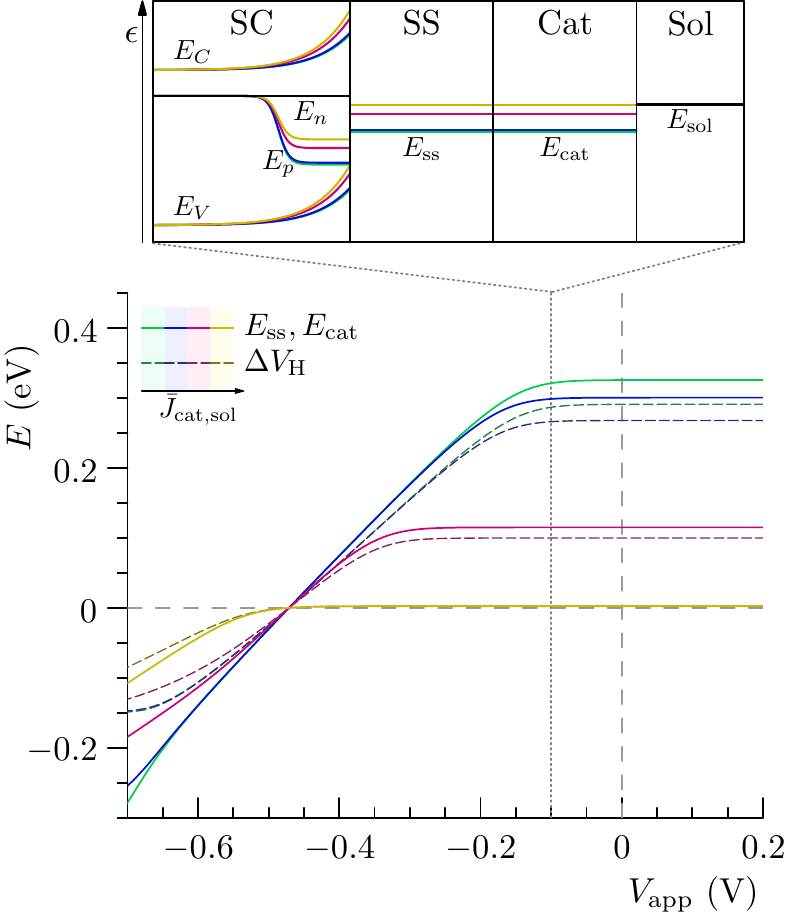} \caption{$\ess$, $\ecat$, $\vh$, and $E_p$ for the series surface state--catalyst model [see Fig \ref{jfig2}(a)--(c), Sec \ref{sercat}] } \label{e4fig}\end{center}\end{figure}

First we investigate a ``series'' catalyst, one which transfers charge between the surface states and solution, but does not directly transfer with the semiconductor. Here we assume that $\jo_\sscat$ is large, so that the surface states and catalyst are at quasiequilibrium. This is reflected in the equality of potentials $\ess=\ecat$, as evident in Fig \ref{e4fig}. Because all charge passes through the surface states, whether it passes from there to the solution or through the catalyst, this is surface state-mediated transfer \cite{vanmaekelber97} (Sec \ref{ssmod}).

The results are shown in Figs \ref{jfig2}(a)-(c) and Fig \ref{e4fig}. With a slow catalyst (the green curve), the result is essentially the same as that obtained in the metallic model [Sec \ref{metcat} - compare the green curves in Figs \ref{jfig1}(c), (d) to Figs \ref{jfig2}(a), (c)]. As $\jocs$ is increased, the total current tends toward the limiting curve of high $\joss$ in the metallic model [Fig \ref{jfig1}(c), purple curve].

Since the catalyst is at quasiequilibrium with the surface states, and functions in series with them, the catalyst effectively increases the rate of transfer from surface states to solution, i.e.\ increases the \textit{effective} value of $\joss$;
\bq \joss[\text{eff}] = e^{\Delta V_H} J_G s^+ \left(1+\frac{e^{\ecat}}{e^{\esso}}\right)\eq

Note the similarities in $J$ and $J_\text{rec}$ between this model and the metallic model, i.e.\ between Figs \ref{jfig1}(c), (d) and Figs \ref{jfig2}(a), (c). Here the catalyst is taking up some of the load that the surface states with higher values of $\jo_\sssol$ in the metallic model were. Fig \ref{jfig2}(b) shows the current transferred from catalyst to solution. With the lowest value of $\jocs$, the current is all produced by the surface states. In the next (blue) curve, the catalyst is carrying almost half of the total current, the rest provided by the surface states; there is only a modest increase in $J$. As $\jocs$ increases further, however, practically all of the current is carried by the catalyst, and the curve tends toward the ideal photodiode curve, with the barrier height reduced by $\vho$ as discussed in Sec \ref{metcat}.

The maximum attainable $\voc$ for this system is still limited by the equilibrium Helmholtz potential $\vho$. Because the catalyst is effectively only increasing the rate of transfer from surface states to solution, there is no mechanism by which it can alter $\vho$. So while the catalyst can increase the effective exchange current between surface states and solution, thereby increasing the total current, it cannot increase the attainable photovoltage or $\voc$.

\subsubsection{Parallel effect --- compensating for $\vh$}\label{parcat}

\begin{figure}\begin{center}\includegraphics[scale=1]{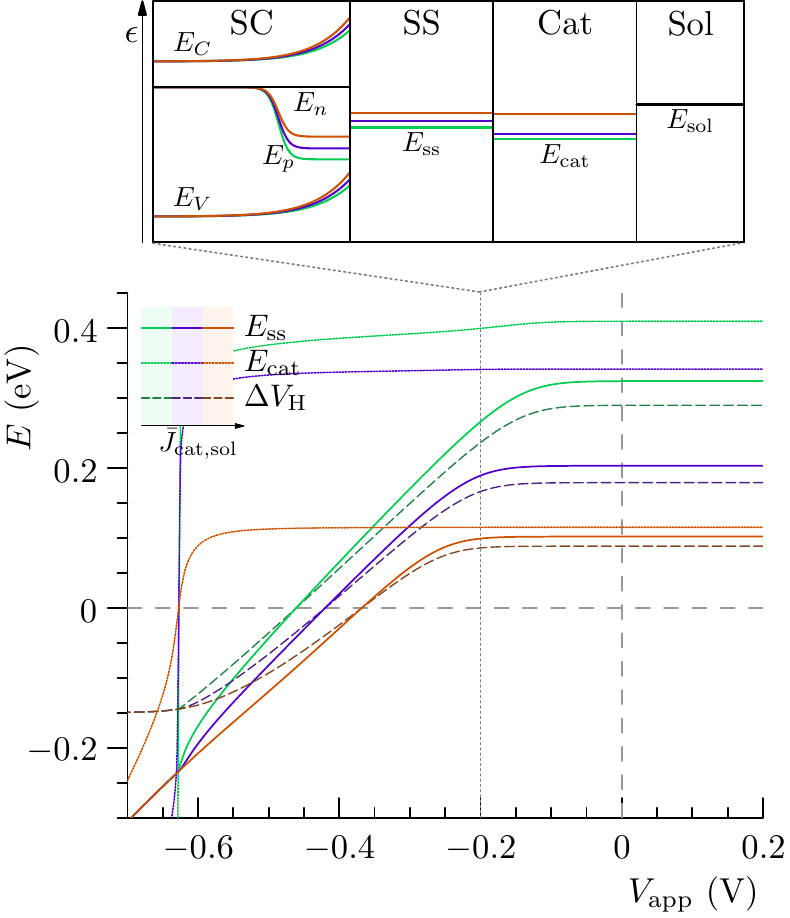} \caption{$\ess$, $\ecat$, $\vh$, and $E_p$ for the parallel surface state--catalyst model [see Fig \ref{jfig2}(d)--(f), Sec \ref{parcat}].} \label{e5fig}\end{center}\end{figure}

We now discuss ``parallel'' effects of the catalyst, in which surface state-catalyst transfer is negligible, but semiconductor-catalyst transfer is comparable to the rate of semiconductor-surface state transfer. In order to clearly demonstrate these effects, we use a slightly faster rate of hole transfer to catalyst than to surface states ($\jo_\vbcat/\jo_\vbss=10^2$), but the electron transfer rate is the same ($\jo_\cbcat=\jo_\cbss$).

The total current $J$ is shown in Fig \ref{jfig2}(d). With a small value of $\jocs$, this mostly behaves like the corresponding (green) curve in the metallic model. However, even with a low value of $\jocs$, a small current is passed through the catalyst [Fig \ref{jfig2}(e)]. Compare the parallel model recombination current Fig \ref{jfig2}(f) to the series model recombination current Fig \ref{jfig2}(c). Even between the green curves (at the lowest $\jo_\catsol$), there is more recombination current in the parallel model. This is because the catalyst is acting as an additional recombination center. This is reflected by the somewhat diminished surface recombination currents, shown in dotted lines in Fig \ref{jfig2}(f). Despite the greater recombination, the small catalyst current [Fig \ref{jfig2}(e)] compensates for this, leading to a net positive current.

Consider the blue and red curves in Fig \ref{jfig2}(d). $\voc$ has been shifted back to the value it had in the absence of $\vho$ [i.e.\ to that found in the screened models, Fig \ref{jfig1}(a)-(b)]. Even though there is still a Helmholtz potential in this system, since the catalyst is moving charge directly from semiconductor to solution and bypassing the surface states (acting in parallel), it circumvents the limitation due to $\vho$, since catalyst-solution transfer is not directly affected by $\vho$ as it is with surface states. Fig \ref{e5fig} shows that $\ecat$ becomes positive near $\vap=-0.63\,\text{V}$, where $\Delta \vh \approx -.15\,\text{V}$. Since $\vho \approx -.15\,\text{V}$ in this system, this is where $\vh\approx 0$, so the presence of the catalyst allows the system to behave as though the reduction in barrier height $\vho$ does not exist, increasing the attainable photovoltage and $\voc$.

\subsubsection{Non-catalytic overlayers}

The use of non-catalytic overlayers, which do not have any apparent redox states capable of transferring to solution, has been found experimentally to increase the performance of photoelectrochemical systems. This type of overlayer would correspond in our model to a system with $\jo_\catsol=0$. The explanation  for the improved performance is usually that these overlayers pacify the surface states, leading to lower surface recombination, or modify the potential distribution at the interface, increasing band bending.

While these are the more plausible physical mechanisms, we note here that in the context of the current model, it is possible to explain an increase in performance with $\jocs=0$ purely kinetically. If a surface-state system is in a regime that is affected by depletion recombination, or is limited by semiconductor-surface state transfer, adding a non-catalytic overlayer can function to increase the transfer of holes to the surface states, decreasing the hole concentration needed to achieve a given current. This can partially counteract the effect of depletion recombination or kinetic limitation and provide a negative shift of the $J(V)$ curve.

\section{Conclusion}

The main conclusions of this work are:

\begin{itemize}
 \item Systems with only ion-permeable catalysts or surface states correspond to our previously defined adaptive, molecular, and metallic catalyst models.
\item Catalysts that act primarily in series with surface states can increase the effective rate of transfer from surface states to solution, leading to an increase in total current
\item Catalysts that act primarily in parallel with surface states can increase the attainable $\voc$ or photovoltage
\item Both series and parellel catalyst effects operate to decrease the Helmholtz potential and surface recombination
\item Both series and parallel catalyst effects operate in tandem in real devices, leading to an increase in current and/or photovoltage, depending on the exchange currents
\end{itemize}

\section{Acknowledgements}
This work was supported by the Department of Energy, Basic Energy Science, Grant DE-SC0014279. F.A.L.L. acknowledges a graduate research fellowship.
\clearpage

\bibliographystyle{unsrt}
\bibliography{scs-refs}

\end{document}